# Synthesis and characterization of FePt/Au core-shell nanoparticles

P. de la Presa[a*], M. Multigner[a], M.P. Morales[b], T. Rueda[a], E. Fernández-Pinel[a] and A. Hernando[a]

[a]*Instituto de Magnetismo Aplicado (UCM-ADIF-CSIC), P.O. Box 155, 28230 Las Rozas, Madrid, Spain*
[b]*Instituto de Ciencia de Materiales de Madrid, CSIC Cantoblanco, 28049 Madrid, Spain*

**Abstract**

In this work, the structural and magnetic properties of the gold-coated FePt nanoparticles synthesized from high-temperature solution phase are presented. The amount of gold was optimized to obtain most of the FePt particles coated. The particle diameter increases from 4 to 10 nm as observed by TEM. The magnetic properties are largely affected by the coating. At low temperature, the coercive field Hc of the coated nanoparticles decreases about three times respect to the uncoated and the blocking temperature reduces to the half. The changes of the magnetic behavior are discussed in terms of the effect of the gold atoms at the FePt core surface.

## 1. Introduction

Surface modification of nanometer sized inorganic core with different inorganic shell to form core/shell type nanostructures has become an important route to functional nanomaterials. Such modification has brought about interesting physical and chemical properties of the nanostructured materials that have shown important technological applications.

The use of magnetic nanoparticles (NPs) is limited in biotechnology by their specific recognition and capture by the macrophages of blood plasma (opsonization process) which depends in great extent on the NPs size and surface. In the case of metallic nanoparticles with high magnetic moments like iron oxides, they have to be protected by coating with a surfactant or an inorganic material because due to their large surfaces they are easily oxidized or otherwise subject to corrosion. In contrast to that, gold nanoparticles are ideal material for biomedical application because of their resistance against oxidation, its biocompatibility and the advances in the thiol chemistry.

In this sense, FePt nanoparticles are promising candidates for biological applications because of the high saturation magnetization and uniformity of the material, which could be achieved by using the synthesis method described by Sun and coworkers[1,2,3]. This material was first proposed for application in high density magnetic storage media but it have opened up new possibilities to exploring further applications of magnetic nanoparticles in biomedicine. Recently, the cytotoxicity of unmodified FePt NPs was found to be not significant during 24 h in brain endothelial cells.[4]

However, the coating by biocompatible materials like gold or silica seems to be the better way to diminish the cytotoxicity grade of the magnetic nanoparticles and, at the same time, to exploit the wide range of surface functionalities of both materials. The studies on coating of magnetite or other ferrite materials are more or less spread out[5,6,] whereas the literature for the effective surface functionalization of FePt, is very sparse.[7]

The aim of this work is to coat FePt nanoparticles by gold and to study the structural characteristics and the changes of the magnetic properties.

## 2. Results and Discussion

The 4 nm FePt nanoparticles, with composition near to 50-50%, were synthesized by the combination of reduction of Pt(acac)$_2$ and decomposition of Fe(CO)$_5$ in octyl ether solvent.[1] These 4 nm FePt nanoparticles were then used as seeds: 5 ml of FePt nanoparticles dispersed in hexane and 30 ml phenyl ether were mixed and stirred under a flow of N$_2$. The mixture was heated at 100 °C for 20 min to remove hexane, and then cooled down to 80º C. After that, 80 mg gold acetate dissolved in 5 ml ethanol, 0.32 ml oleic acid and 0.34 ml oleylamine were added. Under a blanket of nitrogen, the mixture was heated to reflux (265 °C) for another 30 min. The black mixture was cooled to room temperature by removing the heat source. Ethanol (40 ml) was added to the mixture and a black material was precipitated and separated via centrifugation. The black product was dissolved in hexane (25 ml) in the presence of oleic acid (0.05 ml) and oleylamine (0.05 ml). There was roughly 26 %.wt of FePt.

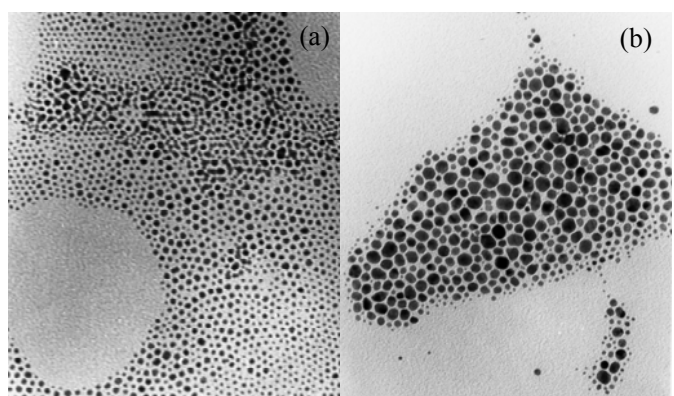

Fig. 1: TEM image of (a) FePt and (b) FePt/Au core-shell NPs.

Particle size was determined from TEM micrographs in a 200 keV JEOL-2000 FXII microscope. For the observation of the sample in the microscope, the particles were dispersed in hexane and a drop of the suspension was placed onto a copper grid covered by a carbon film. The mean particle size was calculated by counting more than 100 particles.

FePt NPs are shown in Fig. 1(a). They are rather spherical in shape and have a mean diameter of 4 nm with a narrow distribution size (standard deviation smaller than 20%). However, gold coated NPs present an irregular shape and mean diameter of around 10 nm with a wide size distribution. The important changes in the aspects of the particles, i.e. irregular shape and increase in size, indicate that a coating layer of gold 3 nm thick has been successfully grown on the FePt particles (Fig. 1(b)).

The coated and uncoated NPs have been magnetically characterized by mean of a Quantum Design SQUID magnetometer. The magnetic characterization consists in magnetization cycles at 5 K and 300 K and maximum applied field of 5 T and zero-field-cooled (ZFC) and field-cooled (FC) curves at 1000 Oe (see Fig. 2). At room temperature, the FePt NPs shows superparamagnetic behavior whereas in the gold-coated NPs the superparamagnetic component has a minor contribution, although it is still present (Fig. 2(b)). At 5 K the magnetic moments of the gold coated FePt NPs saturate at 2 T; nevertheless, for the bare NPs the saturation is not reached even at 5 T. The coercive fields of the samples differ considerably among them at low temperature: the FePt NPs have a coercive field of 1350 Oe whereas coated NPs have a coercivity of about 400 Oe (see Fig. 2(a)), showing a noticeable decrease. At room temperature, the coated NPs have a coercivity of 20 Oe. The ZFC-FC curves show maxima that are characteristic for nanoparticles, these maxima are at 25 K and 12 K for uncoated and the coated NPs, respectively. There is a clear decrease of the blocking temperature for FePt/Au core-shell NPs (see Fig. 3).

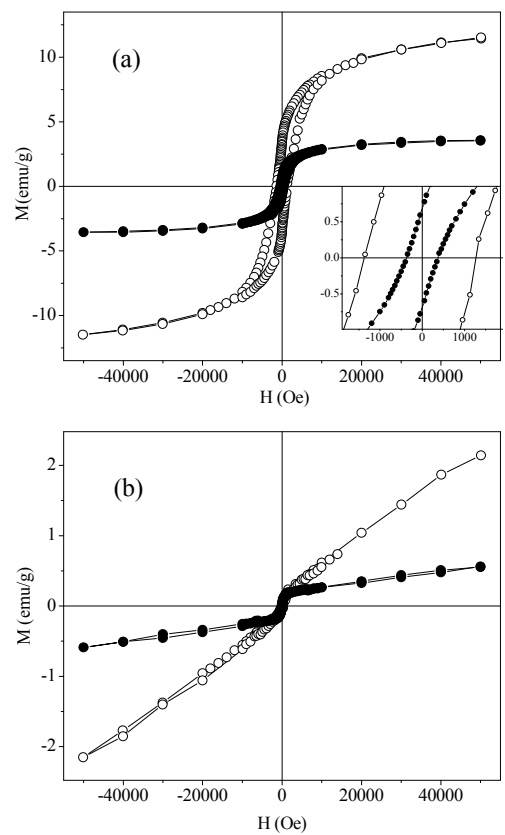

Fig. 2: Magnetization curves of FePt (hollow circles) and FePt/Au core-shell (full circles) at (a) 5 K and (b) 300 K. The inset shows the change in the coercive field. The magnetization is given in emu per gram of sample

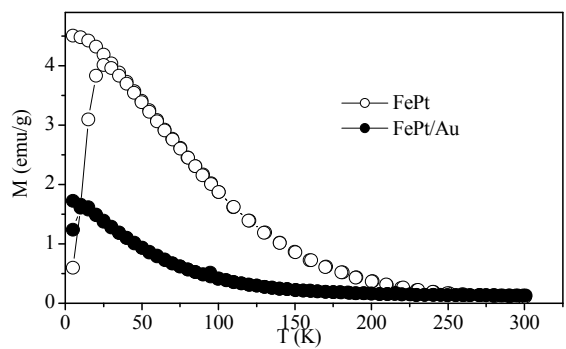

Fig. 3: Zero-field-cool (ZFC)/ field-cool (FC) curves FePt and FePt/Au core-shell nanoparticles The units are emu per gram of sample

The reduction in the blocking temperatures observed in PtFe/$SiO_2$ and $Fe_3O_4$/Au core-shell NPs (Refs. 6,7) have been attributed to the increase in the inter-particle separation (due to the shell) that reduces the magnetic dipole-dipole interaction. In contrast to gold-coated FePt NPs, the coercive field of the gold-coated $Fe_3O_4$ is higher than the uncoated NPs, this increase in the coercivity is attributed to the larger size of the coated NPs and consequently to the less effective coupling of the magnetic dipole moments[6].

However, in the case of FePt/Au core-shell NPs, the blocking temperature as well as the coercive field decreases respect to the bare NPs. The blocking temperature and the coercive field are related to the magnetic anisotropy, as it well known. Since the FePt NPs are about 4 nm in size, we can assume they are monodomain, therefore the blocking temperature is proportional to the anisotropy energy $T_B = KV/25k_B$, with K the magnetic anisotropy of the compound, V the volume of the NP and $k_B$ the Boltzman constant. On the other hand, the coercive field can be approximated to $H_c \approx 2K/M_s$ bellow the blocking temperature, with $M_s$ the saturation moment. Since there is no change in the volume of FePt in the coated NPs, then the reduction of the coercive field and the decrease of the blocking temperature in the gold-coated NPs reflect the decrease of the magnetic anisotropy.

Magnetic anisotropy rises from the combination of the spin-orbit coupling in the magnetic atom and the lack of spherical symmetry in the electric charge distribution of its environment. Therefore, in NPs, the magnetic anisotropy must be an average of bulk and surface anisotropy. It has been recently shown (see Ref. 8) that in the case of $Fe_3O_4$ coated with the surfactant oleic acid the surface anisotropy is reduced due to the covalent bond of the organic molecules to the surface atoms. Although the interaction between FePt and Au atoms at the interface is not known, these results induce to think that the presence of the gold atoms lead to a reduction of the surface anisotropy, which explain the reduction of the blocking temperature and the coercivity for the coated NPs.

The interest of these gold-coated NPs lies in their potential biological applications. The fact the FePt NPs, which are themselves interesting for biological applications, can be coated by gold diminishing the cytotoxicity of Fe in blood is an interesting result. The gold coating allows also easy conjugation of DNA, proteins and diverse biomolecules. However, the chief advantages of the gold coated FePt NPs lie in the metallic coat and in the occurrence of a coercive field (about 20 Oe) at room temperature. Since gold is a conducting metal, allows inductive heating[9] which could be added to the heating power by hysteresis losses, thus resulting in a promising material for hyperthermia treatments. Quantitive heating power measurements will be soon performed.

In summary, we have synthesized gold-coated FePt NPs with a gold layer of 3 nm. The magnetic results show a decrease of the blocking temperature as well as of the coercivity with the coat due probably to the reduction of the surface anisotropy. Currently, we are extending these studies to investigate the influence of the gold atoms on the FePt surface atoms.

## 3. Acknowledgment

Financial support from the Comunidad Autónoma de Madrid under Project No. S-0505/MAT/0194 is acknowledged.